\documentclass{article}
\usepackage{amssymb}
\usepackage{amsmath}
\usepackage{graphicx}

\input{tcilatex}

\begin{document}

\title{Green function for the Poisson equation in a general case of
astrophysical interest.}
\author{Lipovka A.A., Meza A. \\
\textit{Department of Research for Physics, Sonora University, }\\
\textit{83000, Hermosillo}, \textit{Sonora, M\'{e}xico}}
\maketitle

\begin{abstract}
In present paper we suggest exact solution of the Poisson problem which
appears in frequently addressed applications regarding calculation of the
gravitational potential of spiral galaxies. We suggest an analytical
solution for the problem in cylindrical coordinates by using the finite
integral transform technique. The final solution is presented as expansion
on the eigenfunctions of the corresponding Sturm-Liouville problem. Green
function of the problem is constructed.
\end{abstract}

\bigskip 


Pacs numbers: 04.60.-m, 04.65.+e, 11.15.-q, 11.30.Ly

\section{\noindent Introduction}

In a wide variety of astrophysical applications there are a great number of
extremely important problems which can be reduced to the Poisson's problem
in cylindrical coordinates. These problems appear in the modeling of the
galaxies, accretion discs, when we are interested in the reconstruction of
the gravitational potential of the object under consideration by using an
observational data. As an example the method of determining of the spiral
galaxies thickness by solution of the Poisson equation should be mentioned $%
\left[ 1,2\right] $. Unfortunately all these solutions deal with very
particular cases of the density distribution function and for this reason
have some restrictions in applying these models to real galaxies. The
mentioned model of Peng $\left[ 1,2\right] $, for example, is based on the
Parenago's density distribution along the z-direction $\rho _{h}(z)\sim \exp
(-\alpha |z|)$ (they suppose that the density distribution function can be
factorized, i.e. $\rho (r,\phi ,z)=\rho _{\sigma }(r,\phi )\rho _{h}(z)$).
But this function can not be considered as satisfactory choice because 1) it
does not have a defined derivative on the galaxy plane $z=0$ and 2) it was
obtained from the barometric formula which suppose (in thermalized case) the
presence of a constant, solid gravitational mass for $z<0$. As one can see
it is not quite right in the case of a galaxy, where the gravitational
potential in a particular point is defined by the density distribution which
in turn should not be written from simple barometric relation. For this
reasons the model mentioned above needs to be generalized.

Another important application which leads to the problem under consideration
is the calculation of the gravitational potential and field of velocities of
stars in the presence of so-called dark matter (DM). See for example $\left[
3\right] $ and references therein. By taking into account a great impact
that the phenomena of DM produces to the modern theoretical physics, the
need for the accurate analytical solution of the Poisson's equation for
gravitational potential of thin oblate ellipsoid, characterized by more
sophisticated density distribution function becomes clear.

Up to now the Poisson's problem solutions were suggested for rather small
class of the density distribution functions. The purpose of present paper is
to fill this gap and suggest the Green function for the Poisson's problem
for the most general case of an arbitrary density distribution function.

On the one hand if we are interested in the DM distribution, the final
solution is approximately no depends on the spiral structure of the galaxy,
so the angular variable can be omitted in first approximation as
insignificant $\left[ 3\right] $ and on the other hand 2D solution is
important because it appears in many applications and allow simple usage,
analysis and interpretation. By taking into account all mentioned above, we
consider 2D problem because of its great importance.

In present paper we solve the 2D Poisson's problem in cylindric coordinates
by using the finite integral transform technique (FITT) $\left[ 4,5,6,7,8,9%
\right] $ first time developed by Grinberg $\left[ 4\right] $ (in this paper
we will refer this method as Grinberg's method). The final solution is
presented as expansion on the eigenfunctions of the corresponding
Sturm-Liouville problem. Green function of the problem is constructed.

\noindent 

\section{Gravitational Potential of Spiral Galaxies}

Consider a symmetrical, flat, spinning, disc-like oblate object
characterized by the density distribution $\rho (r,\phi ,z)$. In this case
the equation of Poisson $\nabla ^{2}u=\varkappa \rho $ in cylindrical
coordinates for gravitational potential $u$ is:%
\begin{equation}
\frac{1}{r}\frac{\partial }{\partial r}\left( r\frac{\partial u}{\partial r}%
\right) +\frac{1}{r^{2}}\frac{\partial ^{2}u}{\partial \phi ^{2}}+\frac{%
\partial ^{2}u}{\partial z^{2}}=\varkappa \rho (r,\phi ,z).  \tag{1}
\end{equation}

As it was mentioned above, the spiral structure of the galaxy is
approximately insignificant in the problems mentioned above (variation of
the stars velocity due to the presence of the spiral structure is more than
order of magnitude smaller if compared with its unperturbed value $\left[ 3%
\right] $) and for this reason here we can consider only two variables $r$
and $z$, by setting $\phi =conts$. In this case the equation of Poisson
becomes:%
\begin{equation}
\frac{1}{r}\frac{\partial }{\partial r}\left( r\frac{\partial u}{\partial r}%
\right) +\frac{\partial ^{2}u}{\partial z^{2}}=\varkappa \rho (r,z)\text{ \ ,%
}  \tag{2}
\end{equation}

where $\rho (r,z)$ is the surface density of the matter.

This equation also should be supplied by following boundary conditions:%
\begin{equation}
\frac{\partial u}{\partial r}\underset{r=0}{\mid }=0\text{ \ , \ }\frac{%
\partial u}{\partial r}\underset{r=a}{\mid }=-\frac{V_{0}^{2}}{a}g(z)\text{
\ ,}  \tag{3}
\end{equation}%
\begin{equation}
\frac{\partial u}{\partial z}\underset{z=0}{\mid }=0\text{ \ , \ }u\underset{%
z=\infty }{\mid }<\infty \text{ \ ,}  \tag{4}
\end{equation}

where $V_{0}$\ is the experimentally measured velocity of stars located at
distance $a$ from center of galaxy. To resolve the problem (2),(3),(4), the
method of Grinberg (FITT method) $\left[ 4,5,6,7,8,9\right] $ is applied. In
consequence with the technic we should firstly define and resolve the
Sturm-Liouville problem (SLP) to obtain the complete set of eigenfunctions
in which the final solution can be expanded. The corresponding homogeneous
equation reads:

\begin{equation}
\frac{1}{r}\frac{\partial }{\partial r}\left( r\frac{\partial u}{\partial r}%
\right) =-\frac{\partial ^{2}u}{\partial z^{2}}=-\lambda \text{ \ .}  \tag{5}
\end{equation}

Let $u(r,z)=R(r)Z(z)$, in this case we obtain corresponding SLP for $R(r)$
which actually is the Bessel equation:

\begin{equation}
(rR^{\prime })^{\prime }+\lambda rR=0,  \tag{6}
\end{equation}

completed by the homogeneous boundary conditions:

\begin{equation}
R^{\prime }\underset{r=0}{\mid }=0\text{ \ ; \ }R^{\prime }\underset{r=a}{%
\mid }=0\text{ \ .}  \tag{7}
\end{equation}

We note here that the SLP based on the $Z(z)$ is not regular one and for
this reason the problem (6),(7) should be favored. As one can see the
equation (6) and boundary conditions (7) do form the SLP. For zeroth
eigenvalue it has solution:

\begin{equation}
\lambda _{0}=0\text{ \ ; \ }R_{0}=1\text{ \ .}  \tag{8}
\end{equation}

In the case when the eigenvalue $\lambda \neq 0$, we obtain the equation of
Bessel, which has the radial solution for this problem expressed with
functions of Bessel:%
\begin{equation}
\lambda _{n}=(\frac{\gamma _{n}}{a})^{2}\text{ \ ; \ }R_{n}=J_{0}(\sqrt{%
\lambda _{n}}r),  \tag{9}
\end{equation}

where the roots $\gamma _{n}=(\sqrt{\lambda _{n}}a)$ satisfy the following
transcendent equation:%
\begin{equation}
J_{1}(\gamma _{n})=0\text{ \ .}  \tag{10}
\end{equation}

For reader's convenience we also write here the absolute values of the
eigenfunctions (we will use them to construct the final solution of the
problem): 
\begin{equation}
\left\Vert R_{n}\right\Vert ^{2}=\int_{0}^{a}R_{n}R_{n}rdr\text{ \ .} 
\tag{11}
\end{equation}

For zeroth eigenvalue $n=0$\ we have

\begin{equation}
\left\Vert R_{0}\right\Vert ^{2}=\frac{a^{2}}{2}\text{ \ ,}  \tag{12}
\end{equation}

and for $n\neq 0$\ this value is 
\begin{equation}
\left\Vert R_{n}\right\Vert ^{2}=\frac{a^{2}}{2}J_{0}^{2}(\gamma _{n}). 
\tag{13}
\end{equation}

It is well known that the eigenfunctions of problem of Sturm-Liouville
(6),(7) form a complete set of functions in Hilbert space. For this reason
if the function $u(r,z)$ comply the Dirichlet condition within interval $%
\left[ 0,\alpha \right] $, it can be expanded into the series of Dini:%
\begin{equation}
u(r,z)=\dsum\limits_{u=0}^{\infty
}C_{n}(z)R_{n}(r)=C_{0}(z)+\dsum\limits_{u=0}^{\infty }C_{n}(z)J_{0}(\sqrt{%
\lambda _{n}}r),  \tag{14}
\end{equation}

where%
\begin{equation}
C_{n}(z)=\frac{\bar{u}_{n}(z)}{\left\Vert R_{n}\right\Vert _{r}^{2}}, 
\tag{15}
\end{equation}

and the transformed potential function $\bar{u}_{n}(z)$ is given by integral%
\begin{equation}
\bar{u}_{n}(z)=\int_{0}^{a}u(r,z)R_{n}(r)rdr.  \tag{16}
\end{equation}

To find the transformed potential function $\bar{u}_{n}(z)$ , in consequence
with the Grinberg's method, we should transform initial equation (2) and the
boundary conditions (4). Transformed equations can be easily obtained by
using the boundary conditions (3) and (7). They can be written as follows:%
\begin{equation}
\frac{d^{2}\bar{u}_{n}(z)}{dz^{2}}-\lambda _{n}\bar{u}_{n}(z)=F_{n}(z)\text{
\ ,}  \tag{17}
\end{equation}

where

\begin{equation}
F_{n}(z)=\varkappa \rho _{n}(z)+V_{0}^{2}J_{0}(\gamma _{n})g(z)\text{ \ ,} 
\tag{18}
\end{equation}

and the transformed surface density\ $\rho _{n}(z)$ is determined by relation%
\begin{equation}
\rho _{n}(z)=\int_{0}^{a}\rho (r,z)R_{n}rdr\text{ \ .}  \tag{19}
\end{equation}

Transformed boundary conditions are:

\begin{equation}
\bar{u}_{n}\underset{z=\infty }{\mid }<\infty \text{ \ ; \ }\frac{\partial 
\bar{u}_{n}}{\partial z}\underset{z=0}{\mid }=0\text{ \ .}  \tag{20}
\end{equation}

The equations (17) with the boundary conditions (20) form the problem to
obtain the transformed potential function $\bar{u}_{n}(z)$.

Now we should consider two particular cases. The first one corresponds to
the zeroth eigenvalue, when $n=0$, $\lambda _{0}=0$ and $R_{0}=1$. In this
case the equation (17) become:%
\begin{equation}
\bar{u}_{0}^{\prime \prime }(z)=F_{0}(z).  \tag{21}
\end{equation}

By taking into account the symmetry of the problem in respect to the $z$
variable, one can write the solution of the equation (21) which satisfy the
boundary conditions (20): 
\begin{eqnarray}
\bar{u}_{0}(z) &=&4\int_{0}^{z}\int_{0}^{z^{\prime }}F_{0}(z^{\prime \prime
})dz^{\prime \prime }dz^{\prime }+B_{0}=  \TCItag{22} \\
&=&4\varkappa \int_{0}^{z}\int_{0}^{z^{\prime }}\int_{0}^{a}\rho
(r,z^{\prime \prime })rdrdz^{\prime \prime }dz^{\prime
}+4V_{0}^{2}\int_{0}^{z}\int_{0}^{z^{\prime }}g(z^{\prime \prime
})dz^{\prime \prime }dz^{\prime }+B_{0}\text{ \ .}
\end{eqnarray}

Consider now the case when $n\neq 0$ i.e. $\lambda _{n}\neq 0$. Solution of
the equation (17) can be written as%
\begin{equation}
\bar{u}_{n}(z)=C_{1}(z)e^{-\sqrt{\lambda _{n}}z}+C_{2}(z)e^{\sqrt{\lambda
_{n}}z},  \tag{23}
\end{equation}

where $e^{\sqrt{\lambda }z}$ and $e^{-\sqrt{\lambda }z}$ are general
solutions of the corresponding homogeneous equations

\begin{equation}
\frac{d^{2}\bar{u}_{n}(z)}{dz^{2}}-\lambda _{n}\bar{u}_{n}(z)=0,  \tag{24}
\end{equation}

and coefficients $C_{1}(z)$ and $C_{2}(z)$\ are some functions to be
determined. The functions $C_{1}(z)$ and $C_{2}(z)$ can be obtained
immediately:%
\begin{equation}
C_{1}(z)=-\frac{1}{\sqrt{\lambda _{n}}}\int_{0}^{z}F_{n}(z^{\prime })e^{%
\sqrt{\lambda _{n}}z^{\prime }}dz^{\prime }  \tag{25}
\end{equation}

and%
\begin{equation}
C_{2}(z)=\frac{1}{\sqrt{\lambda _{n}}}\int_{0}^{z}F_{n}(z^{\prime })e^{-%
\sqrt{\lambda _{n}}z^{\prime }}dz^{\prime }.  \tag{26}
\end{equation}

So, the solution of the problem (17),(20) for the case $\lambda _{n}\neq 0$
i.e. $n\neq 0$ can be written as:

\begin{equation}
\bar{u}_{n}(z)=\frac{1}{\sqrt{\lambda _{n}}}\left[ e^{\sqrt{\lambda _{n}}%
z}\int_{0}^{z}F_{n}(z^{\prime })e^{-\sqrt{\lambda _{n}}z^{\prime
}}dz^{\prime }-e^{-\sqrt{\lambda _{n}}z}\int_{0}^{z}F_{n}(z^{\prime })e^{%
\sqrt{\lambda _{n}}z^{\prime }}dz^{\prime }\right] ,.  \tag{27}
\end{equation}

where $F_{n}(z^{\prime })$ is given by (18), and

\begin{equation}
\rho _{n}(z)=\int_{0}^{a}\rho (r,z)R_{n}(r)rdr\text{ \ .}  \tag{28}
\end{equation}

Now we are ready to construct the Green function and write the final
expression for the gravitational potential $u(r,z)$. The solution of the
problem under consideration described by equation (2) and boundary
conditions (3) and (4), can be written as follow:%
\begin{equation}
u(r,z)=\bar{u}_{0}(r,z)\frac{R_{0}}{\left\Vert R_{0}\right\Vert _{r}^{2}}%
+\dsum\limits_{n=1}^{\infty }\bar{u}_{n}(r,z)\frac{R_{n}(r)}{\left\Vert
R_{n}(r)\right\Vert _{r}^{2}}.  \tag{29}
\end{equation}

By substituting (8,9,12,13,22,27) into (29) we obtain final solution of our
problem:

\begin{equation}
u(r,z)=\frac{8}{a^{2}}\int_{0}^{z}\int_{0}^{z^{\prime }}F_{0}(z^{\prime
\prime })dz^{\prime \prime }dz^{\prime }+\frac{2B_{0}}{a^{2}}+\frac{2}{a^{2}}%
\dsum\limits_{n=1}^{\infty }\bar{u}_{n}(z)\frac{J_{0}(\sqrt{\lambda _{n}}r)}{%
J_{0}^{2}(\gamma _{n})}.  \tag{30}
\end{equation}

where $\bar{u}_{n}(r,z)$\ is given by the relation (27).

Expression (30) describes the gravitational potential for a spiral galaxy in
the case of an arbitrary distribution of the density function $\rho (r,z)$.

\section{Discussion}

Let us apply the final result (30) to some important particular cases of the
widely used density functions $\rho (r,z)$.

Suppose that function $\rho (r,z)$\ can be factorized $\rho (r,z)=\rho
_{r}(r)\rho _{h}(z)$ , and $\rho _{h}(z)=\beta \exp (-\beta |z|)$ is the
Parenago's density distribution along the z-direction, where $\beta =1/z_{0}$%
\ , and $z_{0}$\ is the half depth of the disk.

In this case the integrations \ over $z$ in (27) and (30) can be carried out
analytically and we obtain:

\begin{eqnarray}
u(r,z) &=&\frac{8}{a^{2}}\left\{ \frac{\varkappa \rho _{r0}(a)}{\beta }\left[
e^{-\beta z}-1+\beta z\right] +V_{0}^{2}z^{2}\right\} +\frac{2B_{0}}{a^{2}}+
\TCItag{31} \\
&&+\frac{2}{a^{2}}\dsum\limits_{n=1}^{\infty }\frac{1}{\sqrt{\lambda _{n}}}%
[e^{\sqrt{\lambda _{n}}z}\left\{ \frac{\beta \varkappa \rho
_{rn}(a)(e^{-(\beta +\sqrt{\lambda _{n}})z}-1)}{(-\beta -\sqrt{\lambda _{n}})%
}-\frac{(e^{-\sqrt{\lambda _{n}}z}-1)V_{0}^{2}J_{0}(\gamma _{n})}{\sqrt{%
\lambda _{n}}}\right\} -  \notag \\
&&e^{-\sqrt{\lambda _{n}}z}\left\{ \frac{\beta \varkappa \rho
_{rn}(a)(e^{-(\beta -\sqrt{\lambda _{n}})z}-1)}{(-\beta +\sqrt{\lambda _{n}})%
}+\frac{(e^{\sqrt{\lambda _{n}}z}-1)V_{0}^{2}J_{0}(\gamma _{n})}{\sqrt{%
\lambda _{n}}}\right\} ]\frac{J_{0}(\sqrt{\lambda _{n}}r)}{J_{0}^{2}(\gamma
_{n})},
\end{eqnarray}

where

\begin{equation}
\rho _{r0}(a)=\int_{0}^{a}\rho _{r}(r^{\prime })r^{\prime }dr^{\prime }, 
\tag{32}
\end{equation}

\begin{equation}
\rho _{rn}(a)=\int_{0}^{a}\rho _{r}(r^{\prime })J_{0}(\gamma _{n}\frac{%
r^{\prime }}{a})r^{\prime }dr^{\prime }\text{ \ .}  \tag{33}
\end{equation}

Expressions (31), (32) and (33) suggest solution of the Poisson problem
written in cylindrical coordinates for a particular case of the Parenago's
density distribution along the z-direction.

First integration over $r^{\prime }$ in (31) can be carried out analytically
in some commonly used cases of the density distribution function $\rho (r)$.

a) Plummer $\left[ 10\right] ,\left[ 11\right] $ distribution function

\begin{equation}
\rho (r)=\frac{3M}{4\pi b^{2}}\left( 1+\frac{r^{2}}{b^{2}}\right) ^{-5/2}%
\text{ \ ,}  \tag{34}
\end{equation}

where $b$ is half - depth of the galaxy. Integrating (32) we obtain:

\begin{equation}
\rho _{r0}(a)=\frac{M}{4\pi }\left[ 1-\left( 1+\frac{a^{2}}{b^{2}}\right)
^{-3/2}\right]  \tag{35}
\end{equation}

b) Plummer-Kuzmin model, also known as \textquotedblleft Toomre's model
1.\textquotedblright\ $\left[ 11\right] $ is defined by the distribution
function

\begin{equation}
\rho (r)=\frac{Mb}{2\pi (r^{2}+a^{2})^{3/2}}\text{ \ ,}  \tag{36}
\end{equation}

where $a$ is truncation radius. In this case the integral (32) became

\begin{equation}
\rho _{r0}(a)=\frac{Mb}{2\pi a}\frac{\sqrt{2}-1}{\sqrt{2}}.  \tag{37}
\end{equation}

These relations suggest solutions of the problem for some important
particular cases of density distribution.

\section{\protect\bigskip Conclusions}

In present paper we consider the Poisson problem in cylindrical coordinates,
which often arises in calculation of the gravitational potential of spiral
galaxies.

By the finite integral transform technique, an exact analytic solution of
the problem is obtained for an arbitrary mass distribution function.

Green function of the problem is written out. As an example, solutions for
widely used in astrophysics density functions, namely the functions of
Parenago, Plummer and Plummer-Kuzmin, are suggested. The obtained
expressions make it possible to avoid or reduce the cumbersome numerical
calculations and allow a clear interpretation of the results.

\textbf{Bibliography}

\begin{enumerate}
\item Peng Qui-he. \textit{A method of determining the thickness of spiral
galaxies by solution of the three-dimensional Poisson equation.} (1988)
Astronomy and Astrophysics (ISSN 0004-6361), vol. 206, no. 1, Nov. 1988, p.
18-26.\qquad 

\item Peng Q., Li X., Su H., Huang K., Huang J. \textit{Integral of
Poisson's equations for finite thickness disks and effect of thickness on
density waves.} (1978) Scientia Sinica, v.22, pp.925-933.

\item S.S. McGaugh and F. Lelli. Phys.Rev.Lett 2016

\item Grinberg G.A. \textit{Selected problems of mathematical theory of the
electric and magnetic fenomena}. Academy of sciences USSR 1948.

\item M.D. Mikhailov, M.N. Ozisik, \textit{Unified Analysis and Solutions of
Heat and Mass Diffusion}, John Wiley \& Sons, 1984.

\item R.M. Cotta, \textit{Integral Transforms in Computational Heat and
Fluid Flow}, CRC Press, Boca Raton, FL, 1993.

\item R.M. Cotta and D. Mikhailov. \textit{Integral transform method. }%
(1993) Appl. Math. Modelling, 1993, Vol. 17, March. pp.156-161

\item A.R. Almeida and R.M. Cotta, \textit{On the integral transform
solution of convection-diffusion problems withinunbounded domain}, J FRANKL
I, 336(5), 1999, pp. 821-832

\item J.S. Perez Guerrero, L.C.G. Pimentel, T.H. Skaggs, M.Th. van Genuchten
(2009) \textit{Analytical solution of the advection-diffusion transport
equation using a change-of-variable and integral transform technique}.
International Journal of Heat and Mass Transfer 52 (2009) 3297-3304
doi:10.1016/j.ijheatmasstransfer.2009.02.002

\item H.C. Plummer \textit{On the problem of distribution in globular star
clusters} (1911) MNRAS v.71, p.460

\item J. Binney and S. Tremaine. \textit{Galactic Dynamics}\ (Princeton
Series in Astrophysics), Princeton University Press. Second edition,
January, 27th 2008. ISBN: 9780691130279, 904 pages.
\end{enumerate}

\end{document}